\documentclass{ws-procs975x65}

\begin{document}

\title{Extended $f(R,L_m)$ theories of gravity}

\author{Francisco S.~N.~Lobo}
\address{Centro de Astronomia e Astrof\'{\i}sica da Universidade de Lisboa,\\ Campo Grande, Ed. C8 1749-016 Lisboa, Portugal\\
E-mail: flobo@cii.fc.ul.pt}

\author{Tiberiu Harko}
\address{Department of
Physics and Center for Theoretical and Computational Physics, \\
The University of Hong Kong, Pok Fu Lam Road, Hong Kong\\
E-mail:harko@hkucc.hku.hk}

\begin{abstract}

We consider a maximal extension of the Hilbert-Einstein action and analyze 
several interesting features of the theory. More specifically, the motion is 
non-geodesic and takes place in the presence of an extra force. These models 
could lead to some major differences, as compared to the predictions of 
General Relativity or other modified theories of gravity, in several problems 
of current interest, such as cosmology, gravitational collapse or the 
generation of gravitational waves. Thus, the study of these phenomena may also 
provide some specific signatures and effects, which could distinguish and 
discriminate between the various gravitational models.

\end{abstract}


\bodymatter

\section{Introduction}\label{Intro}

A promising way to explain the late-time accelerated expansion of the Universe 
is to assume that at large scales General Relativity (GR) breaks down, and a 
more general action describes the gravitational field. Thus, in the latter 
context, infra-red modifications to GR have been extensively explored, where 
the consistency of various candidate models have been analysed (see 
\cite{Lobo:2008sg} for a review). Note that the Einstein field equation of GR was first 
derived from an action principle by Hilbert, by adopting a linear function
of the scalar curvature, $R$, in the gravitational Lagrangian density. The
physical motivations for these modifications of gravity were related to the 
possibility of a more realistic representation of the gravitational fields 
near curvature singularities and to create some first order approximation for 
the quantum theory of gravitational fields, and more recently in an attempt to 
explain the late-time cosmic acceleration. In this context, a more general 
modification of the Hilbert-Einstein gravitational Lagrangian density 
involving an arbitrary function of the scalar invariant, $f(R)$, has been
extensively explored in the literature, and recently a maximal extension of 
the Hilbert-Einstein action has been proposed \cite{Harko:2010mv}.

\section{$f(R,L_m)$ gravity}

The action of the maximal extension of the Hilbert-Einstein action is given by 
\cite{Harko:2010mv}
\begin{equation}
S=\int f\left(R,L_m\right) \sqrt{-g}\;d^{4}x~,
\end{equation}
where $f\left(R,L_m\right)$ is an arbitrary function of the Ricci scalar $R$, 
and of the Lagrangian density corresponding to matter, $L_{m}$. The 
energy-momentum tensor of matter is defined as $T_{\mu \nu }=-\frac{2}{\sqrt{-
g}}\frac{\delta \left(\sqrt{-g}L_{m}\right)}{ \delta g^{\mu \nu }}$.
Varying the action with respect to the metric $g^{\mu\nu}$, the gravitational 
field equation of $f\left( R,L_{m}\right)$ gravity is provided by
\begin{eqnarray}\label{field}
f_{R}\left( R,L_{m}\right) R_{\mu \nu }+\left( g_{\mu \nu
}\nabla _{\mu }\nabla^{\mu } -\nabla_{\mu }\nabla _{\nu }\right) f_{R}\left( 
R,L_{m}\right) 
   \nonumber  \\
-\frac{1}{2}\left[ f\left( R,L_{m}\right) -
f_{L_{m}}\left(R,L_{m}\right)L_{m}\right] g_{\mu \nu }=\frac{1}{2}
f_{L_{m}}\left( R,L_{m}\right) T_{\mu \nu }\,.
\end{eqnarray}
For the Hilbert-Einstein Lagrangian, $f( R,L_{m})=R/2\kappa^2+L_{m}$, we 
recover the Einstein field equations of GR, i.e., $R_{\mu \nu }-(1/2)R\,g_{\mu
\nu}=\kappa^2 T_{\mu \nu }$. For $f\left( R,L_{m}\right)=f_{1}(R)+f_{2}
(R)G\left( L_{m}\right) $, where $f_{1}$, $f_{2}$ and $G$ are arbitrary 
functions of the Ricci scalar and of the matter Lagrangian density, 
respectively, we obtain the field equations of modified gravity with an 
arbitrary curvature-matter coupling 
\cite{Harko:2008qz,Bertolami:2007gv,Harko:2010hw}. An interesting application 
was explored in the context of $f(R,T)$ gravity\cite{Harko:2011kv}.

The $f(R,L_m)$ models possess extremely interesting properties. First, the
covariant divergence of the energy-momentum tensor is non-zero, and is given 
by
\begin{eqnarray}
\nabla ^{\mu }T_{\mu \nu }=2\nabla
^{\mu }\ln \left[ f_{L_m}\left(R,L_m\right) \right] \frac{\partial L_{m}}{%
\partial g^{\mu \nu }}\,.  \label{noncons}
\end{eqnarray}
The requirement of the conservation of the energy-momentum tensor of matter, 
$\nabla ^{\mu }T_{\mu \nu }=0$, provides the condition given by $\nabla ^{\mu 
}\ln \left[ f_{L_m}\left(R,L_m\right) \right] \partial L_{m}/
\partial g^{\mu \nu }=0$.

Secondly, the motion of test particles is non-geodesic, and takes place in the 
presence of an extra force. As a specific example, consider the case in which 
matter, assumed to be a perfect thermodynamic fluid, obeys a barotropic 
equation of state, with the thermodynamic pressure $p$ being a function of the 
rest mass density of the matter $\rho $ only, i.e., $p=p(\rho )$, and 
consequently, the matter Lagrangian density, becomes an arbitrary function of 
the energy density $\rho $ only, i.e., $L_{m}=L_{m}\left( \rho \right)$ 
(for more details, we refer the reader to
\cite{Harko:2010mv,Bertolami:2008ab,Bertolami:2008zh}). Thus, the equation of 
motion of a test fluid is given by
$d^2x^\mu /ds^2+\Gamma _{\nu \lambda }^{\mu }u^{\nu}u^{\lambda }=f^{\mu }$,
where the extra-force $f^\mu$ is defined by
\begin{equation}
f^{\mu }=-\nabla _{\nu }\ln \left[  f_{L_m}\left(R,L_m\right) \frac{%
dL_{m}\left( \rho \right) }{d\rho }\right] \left( u^{\mu }u^{\nu
}-g^{\mu \nu }\right) \,.
\end{equation}
Note that $f^\mu$ is perpendicular to the four-velocity, $u^{\mu}$, i.e., 
$f^{\mu }u_{\mu }=0$.

The non-geodesic motion, due to the non-minimal couplings present in the 
model, implies the violation of the equivalence principle, which is highly 
constrained by solar system experimental tests. However, it  has recently been 
argued, from data of the Abell Cluster A586, that the interaction between dark
matter and dark energy implies the violation of the equivalence principle 
\cite{BPL07}. Thus, it is possible to test these models with non-minimal 
couplings in the context of the violation of the equivalence principle. It is 
also important to emphasize that the violation of the equivalence principle is 
also found as a low-energy feature of some compactified versions of
higher-dimensional theories.

In the Newtonian limit of weak gravitational fields \cite{Harko:2010mv}, the 
equation of motion of a test fluid in $f\left(R,L_m\right)$ gravity is given 
by
\begin{eqnarray}
\vec{a}=\vec{a}_{N}+\vec{a}_{H}+\vec{a}_{E}  
=-\nabla \phi -\nabla \frac{dL_{m}\left( \rho \right) }{d\rho }-\nabla U_{E}
    \,,
\end{eqnarray}
where $\vec{a}$ is the total acceleration of the system; $\vec{a}_{N}=-\nabla 
\phi $ is the Newtonian gravitational acceleration; the term 
$\vec{a}_{H}=-\nabla \left[dL_{m}\left( \rho \right) /d\rho \right] $ is 
identified with the hydrodynamic acceleration term in the perfect fluid Euler 
equation. Now, by assuming that in the Newtonian limit the function 
$f_{L_m}\left(R,L_m\right)$ can be represented as
$f_{L_m}\left(R,L_m\right)\approx 1+U\left(R, L_m\right)$,
where $U\left(R,L_m\right)\ll 1$, so that $\vec{a}_{E}$ given by
\begin{equation}
\vec{a}_{E}=-\nabla U_{E}=-\nabla \left[ U\left(R,L_m\right)\frac{dL_{m}\left( 
\rho \right)}{d\rho }\right]\,,
\end{equation}
is a supplementary acceleration induced due to the modification of the action 
of the gravitational field.

\section{Conclusion}

In conclusion, the maximal extensions of GR, namely the $f(R,L_m)$ gravity 
models open the possibility of going beyond the algebraic structure of the 
Hilbert-Einstein action. On the other hand, the field equations of $f(R,L_m)$ 
gravity are equivalent to the field equations of the $f(R)$ model in empty 
space-time, but differ from them, as well as from GR, in the presence of 
matter. Thus, the predictions of $f(R,L_m)$ gravitational models could lead to 
some major differences, as compared to the predictions of standard GR, or 
other generalized gravity models, in several problems of current interest, 
such as cosmology, gravitational collapse or the generation of gravitational 
waves. The study of these phenomena may also provide some specific signatures 
and effects, which could distinguish and discriminate between the various 
gravitational models. In addition to this, in order to explore in more detail 
the connections between the $f(R,L_m)$ theory and the cosmological evolution, 
it is necessary to build some explicit physical models.

\section*{Acknowledgements}
FSNL acknowledges financial support of the Funda\c{c}\~{a}o para a Ci\^{e}ncia 
e Tecnologia through the grants CERN/FP/123615/2011 and CERN/FP/123618/2011.

\end{document}